\documentclass[a4paper,UKenglish]{lipics-v2016}
 
\usepackage{microtype}
\usepackage{mathtools}
\usepackage{float}
\usepackage{cleveref}
\makeatletter
\newcommand*{\rom}[1]{\expandafter\@slowromancap\romannumeral #1@}
\makeatother

\DeclarePairedDelimiter{\ceil}{\lceil}{\rceil}

\title{Computing Minimum Weight Cycles to 
Leverage Mispricings in 
Cryptocurrency Market Networks
}
\titlerunning{Computing Minimum Weight Cycles in Cryptocurrency Market Networks} 
\author[1]{Francesco Bortolussi}
\author[1]{Zeger Hoogeboom}
\author[1,2]{Frank W. Takes}
\affil[1]{LIACS, Department of Computer Science, Leiden University, The Netherlands\\
  \texttt{\{f.bortolussi,z.hoogeboom\}@umail.leidenuniv.nl}, \texttt{ftakes@liacs.nl}}
\affil[2]{CORPNET, University of Amsterdam, The Netherlands}
\authorrunning{F. Bortolussi, Z. Hoogeboom and F.W. Takes} 

\Copyright{Francesco Bortolussi, Zeger Hoogeboom and Frank W. Takes}


\begin{document}

\maketitle


\begin{abstract}
Cryptocurrencies such as Bitcoin and Ethereum have recently gained a lot of popularity, not only as a digital form of currency but also as an investment vehicle.  
Online marketplaces and exchanges allow users across the world to convert between dozens of different cryptocurrencies and regular currencies such as euros or dollars. 
Due to the novelty of this concept, the volatility of these markets and the differences in maturity and usage of particular marketplaces, currency pairs may appear at multiple marketplaces but at different trading prices. 

This paper proposes a novel algorithmic approach to take advantage of these mispricings and capitalize upon the pricing differences that exist between exchanges and currency pairs. 
To do so, we model each combination of a currency and a market as one node in a graph. 
A directed link between two nodes indicates that a conversion between these two currency/market pairs is possible. 
The weight of the link relates to the exchange rate of executing this particular currency exchange. 
To leverage the mispricings, we seek for cycles in the graph such that upon multiplying the weights of the links in the cycle, a value greater than $1$ is found and thus a profit can be made. 
Our goal is to do this efficiently, without exhaustively enumerating all possible cycles in the graph.
Therefore, we convert our data and address the problem in terms of finding minimum weight triangles in graphs with integer weights, for which efficient algorithms can be utilized. 

We experiment with parameter settings (heuristics) related to the conversion of exchange rate data into integer weight values. 
We show that our approach improves upon a reasonable baseline algorithm in terms of computation time.  
Furthermore, using a real-world dataset, we demonstrate how the obtained minimal weight cycles indeed unveil a number of currency exchange cycles that result in a net profit. 
Ultimately, these findings pave the way for fully automated real-time leveraging of cryptocurrency market mispricings. 

 \end{abstract}


\section{Introduction}

This paper deals with so-called \emph{cryptocurrencies}, which are digital forms of currencies that rely on cryptography and a public ledger called a blockchain~\cite{pilkington201611}. 
Cryptocurrencies typically have a name and an abbreviation, and well-known examples are Bitcoin (``BTC''), Ethereum (``ETH'') and Litecoin (``LTC''). 
Online exchanges and marketplaces facilitate the exchange of one currency to another, including real-world currencies like the US Dollar (``USD'') and Euro (``EUR''). 
Currency markets allow users to hold a number of different currencies and exchange from one currency to the other against the rate of that particular exchange market. 
Because cryptocurrencies are rather new, markets are still relatively volatile, and marketplaces have diverse maturity, volume and customer counts. 
As a result, the precise exchange rate from a particular currency to another is not identical across marketplaces. 
Therefore, in theory, making a series of exchanges of one currency to another, ultimately back to the currency one started with, may result in a profit. 

\begin{figure}[!b]
	\centering
	\includegraphics[scale=0.35]{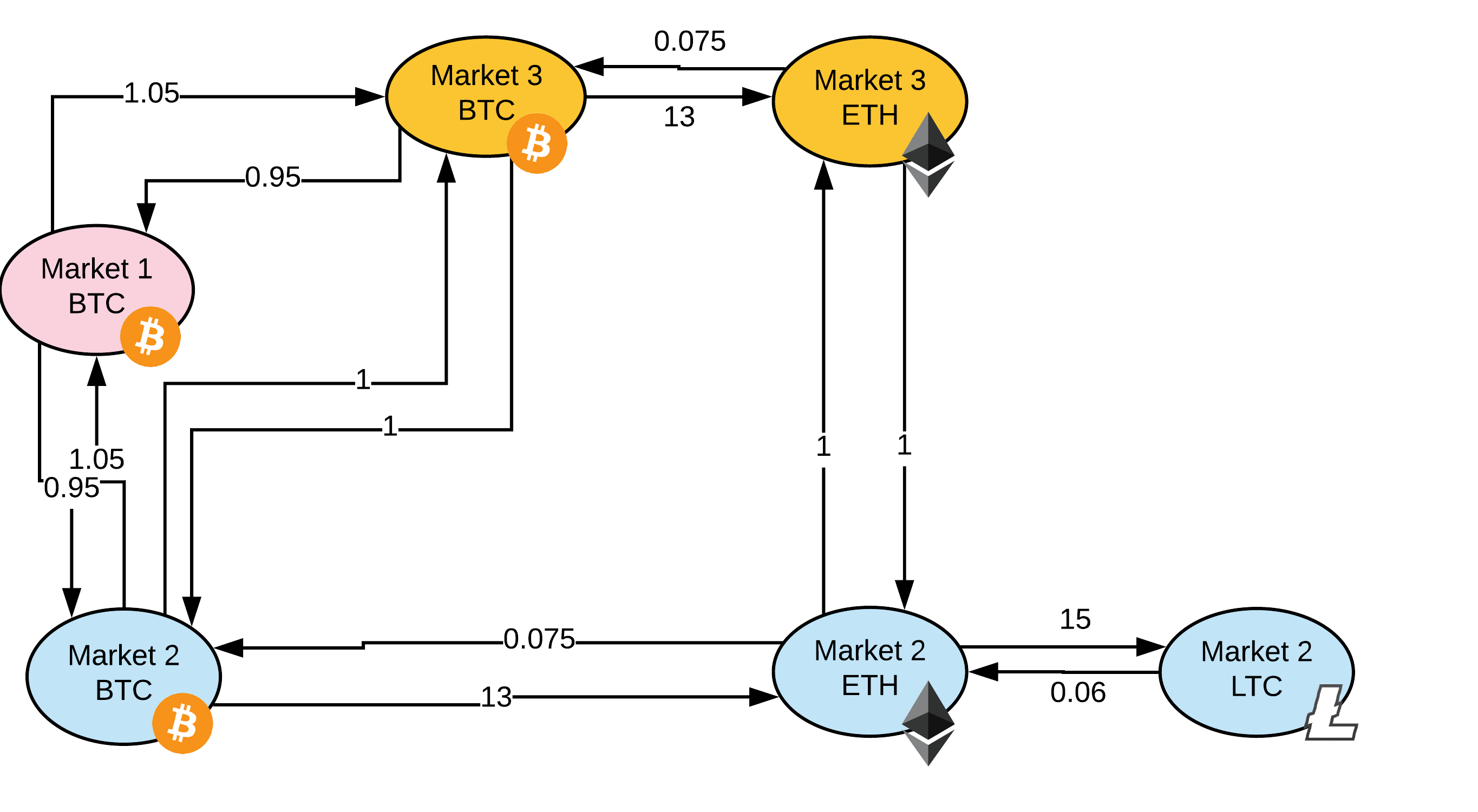}
	\caption{Example of a fictive cryptocurrency market network.}
	\label{fig:example_graph}
\end{figure}

The traditional real-world foreign currency exchange market is a well-known financial market in which by  buying dollars with euros for a given exchange rate, and later selling these dollars for euros, one can potentially make a profit at a particular point in time. 
Here, the focus is on the profit that can be made by taking advantages of the aforementioned pricing differences between markets. 
To do so, we model these cryptocurrencies and their exchange as a directed network or graph, which we call the \emph{cryptocurrency market network}. 
In such a network, the nodes are currencies traded at a particular exchange, and the links denote the exchange rate between these currencies. 
A toy example is given in Figure~\ref{fig:example_graph}. 
Given the network as an underlying model, finding sequences of trades that result in a profit is equivalent to finding cycles that result in a value greater than $1$, when multiplying the weights of the links traversed in the cycle.

For computational reasons discussed later, the abovementioned problem is by means of a transformation of the data translated into the problem of finding minimum weight cycles in $n$-node directed graphs with edge weights. 
Since the edges can be traversed in both directions, the graphs considered in this paper are bidirected; while the links are symmetric, the weights of these edges mainly depend on the direction. 
It was shown in \cite{Roditty2011MinimumAlgorithms} that the minimum weight cycle problem is equivalent to many other graph and matrix problems for which no truly subcubic $(O(n^{3 - \epsilon})$-time for constant $\epsilon$ > 0) algorithms are known. 
It was shown that if there is a truly subcubic algorithm for the minimum weight cycle problem, then many other problems such as the well-known All-Pairs-Shortest-Paths (APSP) problem also have subcubic algorithms. 
However, recent findings show that the minimum weight cycle could actually be computed by finding the minimum weight triangle, which has a significantly lower computation complexity compared to APSP \cite{Roditty2011MinimumAlgorithms}. 
In this paper, we will modify these algorithms such that they can be applied to our cryptocurrency market networks. 

The aim of this paper is to discover whether profitable cycles in these networks exist, and to assess the effectiveness of minimum weight cycle algorithms on this type of data.
Therefore, we do not operate on a marketplace in real time, but instead use a local copy of the state of the cryptocurrency market network. 
This allows us to abstract away from the real-world difficulties of implementing this algorithm in a streaming context with dynamic data and as well as trading volume requirements. 
Regardless, the approach devised in this paper can directly be utilized in a real-world setting, once these implementation are taken care of.

The remainder of the paper is structured as follows.
In \Cref{Problem statement} we give a precise description of the problem as well as the goals we set out to achieve. 
In \Cref{Related Work} there is a description of the previous work related to the topic discussed in the paper. 
In \Cref{Our Approach} the algorithmic approach is presented.
Then, using the data described in \Cref{Datasets}, experiments with our approach are discussed in \Cref{Experiments}. 
Finally, \Cref{Conclusions} concludes the paper and gives suggestions for future work.


\section{Problem statement} \label{Problem statement}

Let $G(V, E, w)$ be a weighted \emph{graph}, where $V$ is the set of nodes, $E$ is the set of edges and $w$ represents the function that outputs the weight value of an edge. 
In particular, $w(u, v)$ or in short $w_e$ is the weight of the edge $e = (u,v) \in E$. 
Only simple graphs were considered. 
A simple graph contains no self loops or parallel edges. 
Unless otherwise specified, $n = |V|$.
A \emph{cycle} is defined as a sequence of $\ell$ nodes $(v_1, \ldots, v_\ell)$ such that, $\forall i < \ell$, $(v_i, v_{i+1}) \in E \wedge (v_\ell, v_1) \in E$. 
We set $w(C)$ and $p(C)$ to be respectively the sum and the product of all edge weights in cycle $C$. 

The connection between currencies and markets is modeled with a graph representation: the nodes are currencies belonging to a certain market while the edges are the exchange rates between two currencies. 
Given an edge $(u, v) \in E$, the edge weight is equal to the amount of currency $v$ one would get for a unit of currency $u$. 
On the other hand, the edge weight of $(v, u) \in E$ is set to $\frac{1}{w(u, v) \epsilon}$ with $0 < \epsilon < 1$. 
This means that the graph is bi-directional, but that because of $\epsilon$, which is set to a value close but not equal to $1$, the cycle of going back and forth between nodes would never result in a profit.
So, the main problem addressed in this paper is as follows:

\begin{quote}
	Given a weighted graph $G$, find a cycle $C$ such that $p(C)$, the product of the weights in this cycle, is maximized.
\end{quote} %

\noindent Here, $p(C) > 1$ represents an overall profit achieved as a consequence of taking advantage of potential mispricings in one or multiple markets. 
Note that in fact, through a transformation algorithm further elaborated on in \Cref{Our Approach}, the objective is actually translated into the following problem statement: 

\begin{quote}
	Given a weighted graph $G'$, find a cycle $C$ such that $w(C)$, the sum of the weights in this cycle, is minimzed.
\end{quote} 

\noindent
Historically, calculating such a \emph{minimum weight cycle} for a directed graph was comparable to calculating the APSP matrix, with respect to computational cost. 
As such, as a naive algorithm for solving this problem, one could use the well-known Floyd-Warshall algorithms~\cite{floyd1962algorithm}  
with some slight modifications: the aforementioned transformation of weights has to be computed, which will be explained in \Cref{Our Approach}. 
Moreover, in the initial adjacency matrix, the distance of every node to itself must be set to $\infty$; this means that the diagonal of the adjacency matrix would be filled with $\infty$ instead of $0$ values. 
After applying the algorithm, the minimum weight cycle can be found by comparing the path of each node to itself.
This baseline algorithm would be significantly slower, with a complexity of $O(n^3)$.
The remainder of this paper focuses on more efficient approaches for finding the minimum weight triangle. 


\section{Related Work} \label{Related Work}

The problem of finding a minimum weight cycle was first studied in \cite{Karp1978ADigraph}. 
It was shown that the problem of finding a minimum weight cycle can be reduced to the problem of finding all pairs shortest paths. 
Given a distance matrix $D$ with empty cells set to $\infty$, which satisfies for every $u, v \in V$, $d[u, v] \leqslant D[u, v]$, the minimum weight cycle has weight $\min_{u,v} D[u, v] + w(u, v)$. 
Using Zwick's APSP algorithm \cite{Zwick2002AllMultiplication}, the minimum weight cycle can be computed in $O(M^{0.681}n^{2.575})$ time. 
However, it was shown in \cite{Roditty2011MinimumAlgorithms} that the minimum weight cycle problem in directed graphs can be reduced to the problem of finding a minimum triangle in an undirected graph which implies that the minimum weight cycle in directed graphs can be computed in $\tilde{O}(Mn^\omega)$ time, improving the original result. 
Here $\omega$ is the exponent of square matrix multiplication with $\omega < 2.373$ \cite{Williams2012MultiplyingCoppersmith-Winograd}. 
In \cite{Roditty2011MinimumAlgorithms}, it was also shown that a minimum weight cycle can be found with the same computational cost in an undirected graph, with a modified version of the algorithm that uses fast matrix multiplication and weights in the interval $[1,M]$.

In \cite{Agarwal2016FindingCycles}, new algorithms were presented to find $k$ multiple simple cycles and simple paths in a graph, in non-decreasing order of their weights. 
After some preprocessing of cost $O(mn + n^2 \log n)$, finding each successive simple shortest cycle in G takes as much as time as the APSP, namely $O(mn+n^ 2 \log \log n)$ time. 
Although finding $k$-shortest cycles is out of the scope of this paper, the results of \cite{Agarwal2016FindingCycles} could also be applied to an extended version of the problem, where more than one cycle is taken into consideration.

In \cite{Roditty2013ApproximatingGirth.}, other studies that were conducted on undirected graphs showed that approximations could be done more efficiently, if one restricts the maximum edge weight to a certain number. 
However, this is not relevant in our case, as we do not have strict bounds on the maximum edge weight. 

For a more generic overview of work on cryptocurrencies, the reader is referred to \cite{tschorsch2016bitcoin}. 
Most other research in this area appears to deal with attempting to understand the price of cryptocurrencies such as Bitcoin~\cite{bouoiyour2015does} or focuses on the topological properties of the network of Bitcoin transactions~\cite{kondor2014rich}.
Yet, the type of cryptocurrency market network data considered in this paper has to the best of our knowledge not yet been investigated. 
In this paper, we utilize the aforementioned findings with respect to minimum weight cycles findings and build on the techniques proposed in \cite{Roditty2011MinimumAlgorithms} to solve our problem of finding the cycle with maximal profit in such a cryptocurrency market network.


\section{Datasets} \label{Datasets}

The dataset that was considered for the experiments reported in this paper was comprised of data gathered from a number of different cryptocurrency markets. 
It focused on both major cryptocurrencies such as Bitcoin and Ethereum as well as a few ``base'' currencies like the US Dollar. 
In total, $110$ different currencies are considered. 
The multiple datasets studied were all in fact snapshots, so ``offline'' versions of the state of the markets at one point in time, in this case from a particular moment in time in January 2018. 
As an example, the average price in US dollar (USD) of $1$ Bitcoin (BTC) was around $11,000$ dollars.

Normally, the exchange rates are constantly changing, and profitable trades are only possible in limited time windows. 
So, latency would be a huge factor in the success or failure of the pipeline; specifically, every trade requires a minimum amount of time dictated by latency. 
In a network, every step would take a minimum amount of time $t$ to transfer the currency. 
This means that, for $\ell$ steps (e.g., a cycle of length $\ell$) in a network (every step is a trade between two currencies), the time required for the whole process to finish is $\ell \cdot t$. 
Throughout this paper we assume that all $\ell$ trades can be executed simultaneously, i.e., within the time window $t$. 
We say this because we assume that there is an infinite amount of currency available to move through the network. 
This is in fact reasonable, as large investors can typically easily manage to maintain balances in a larger number of currencies. 
All in all, it would mean that all trades can be done simultaneously, thus independent of $t$. 
We note that although $\ell$ and $t$ both no longer play a role in the remainder of the paper, a real-world implementation would have to be able to compute the minimum weight cycles within $t$ time, in order to actually execute the trades.
A final assumption is that there are no obstructive fees for sending money and executing trades. 
In the real world, this assumption is very reasonable, as large-scale investors typically have sufficient volume to have drastically lower currency transfer trading fees. 

Table~\ref{tab:description} describes the different datasets used in this paper. 
Apart from finding the minimum weight cycle for the ``Full'' dataset with $110$ currencies and $16$ different markets, we also want to understand whether a profit can be made in one single market, which is why we also perform experiments for three markets individually. 
Finally, it makes sense to see if it matters whether we include less popular cryptocurrencies as opposed to only the major currencies. 
Therefore, we also create a subgraph based on the largest five currencies in terms of market capitalization (a metric of the amount of the size of this currency given today's exchange rate).  

\begin{table}[!t]
	\centering
	\begin{tabular}{p{4cm}ll|p{6.3cm}}
		\textbf{Dataset} &\textbf{|V|} &\textbf{|E|} &\textbf{Description}\\ \hline
		Full graph & 243 & 1718 & The full dataset, a snapshot from 16 markets  \\ \hline
		Only Market 1, all currencies & 7 & 24 & Only the data for Market 1 \\ \hline
		Only Market 2, all currencies & 34 & 130 & Only the data for Market 2 \\ \hline
        Only Market 3, all currencies & 21 & 92 & Only the data for Market 3 \\ \hline
		Five biggest currencies & 45 & 274 & The 5 cryptocurrencies which have the largest market capitalization\\
	\end{tabular} 
	\caption{Overview of the different cryptocurrency market network datasets.}
	\label{tab:description}
\end{table}

Within a market, not every currency could be traded with every other currency, so the network contains a number of denser subgraphs of different sizes, centered around a particular market. 
Along similar lines, not all currencies are traded at every market and as such, not all markets are equally well connected to other markets. 
Indeed, with $110$ currencies and $16$ markets there could in theory be $110 \cdot 16$ nodes in the network, yet it only has $243$.


\section{Approach} \label{Our Approach}

The methodology used to find profitable cycles is comprised of three steps:

\begin{enumerate}
	\item Transformation of the original graph $G$ to $G'$ in order to change the problem from maximization of the products of the weights to minimization of the sum of the weights. 
	\item Finding the minimum weight triangle in a transformed version of the network, $G'$.
	\item Computing the witness matrix to find the exact nodes of the original network that form the corresponding minimum weight cycle. 
\end{enumerate}

\noindent
In the next subsections we describe each one of these three steps.

\subsection{Modifying edge weights of the dataset}
\label{sec:weightmodi}

As the edge weights are modeled as exchange rates, the main goal was to maximize the product of weights in a cycle. 
This conflicts with the problem of finding a minimum weight cycle, as the problem assumes that a sum of weights of a cycle has to be minimized. 
Furthermore, the algorithm from \cite{Roditty2011MinimumAlgorithms} assumes that the edge weights are integers, where the exchange rates are floating point values. 
For these reason, a transformation was applied. 
Given that the algorithm complexity depends on the possible range of weights $M$, the goal was to map the values to an as small as possible range of integers. 
The algorithm to transform the values is described below:

\begin{enumerate}
	\item For each of the edges $e \in E$ of the initial graph, the new weight is set to $w_e^{\rom{1}} = \frac{1}{w_e}$. This is necessary for the transformation from a maximization to a minimization problem. 
	\item Let $min_w$ be the minimum weight value of the dataset: every weight is multiplied by $k = \frac{1}{min_w}$, i.e. $w_e^{\rom{2}} = k \cdot w_e^{\rom{1}}$.
    With this step, all weights are set to a value $w_e^{\rom{2}} \geq 1$. This is important because of the implications of the following steps.
	\item For each edge, the new weights were set to $w_e^{\rom{3}} = \log{w_e^{\rom{2}}}$. This step is necessary to change the minimum of a product to a minimum of a sum. 
	\item Finally, in order to end up with only integer weights, all $w_e^{\rom{3}}$ are multiplied by a constant parameter $c$, called the ``weight multiplier'', and rounded up. 
    Formally, $w_e^{\rom{4}} = \ceil{c w_e^{\rom{3}}}$.  The way the order of the weights is kept consistent, and the dataset only contains integers.
    The weight multiplier value was chosen to preserve the uniqueness of the floating values after rounding them to integer values.
    As an example, the numbers $10.24$ and $10.34$ would have the same integer rounding, if they were not multiplied by $10$ beforehand.
     We experimented with different values of $c$ in \Cref{Experiments}. 
\end{enumerate}

\begin{figure}[!b]
	\centering
     \begin{subfigure}[b]{0.45\textwidth}
         \centering
         \includegraphics[scale=0.16]{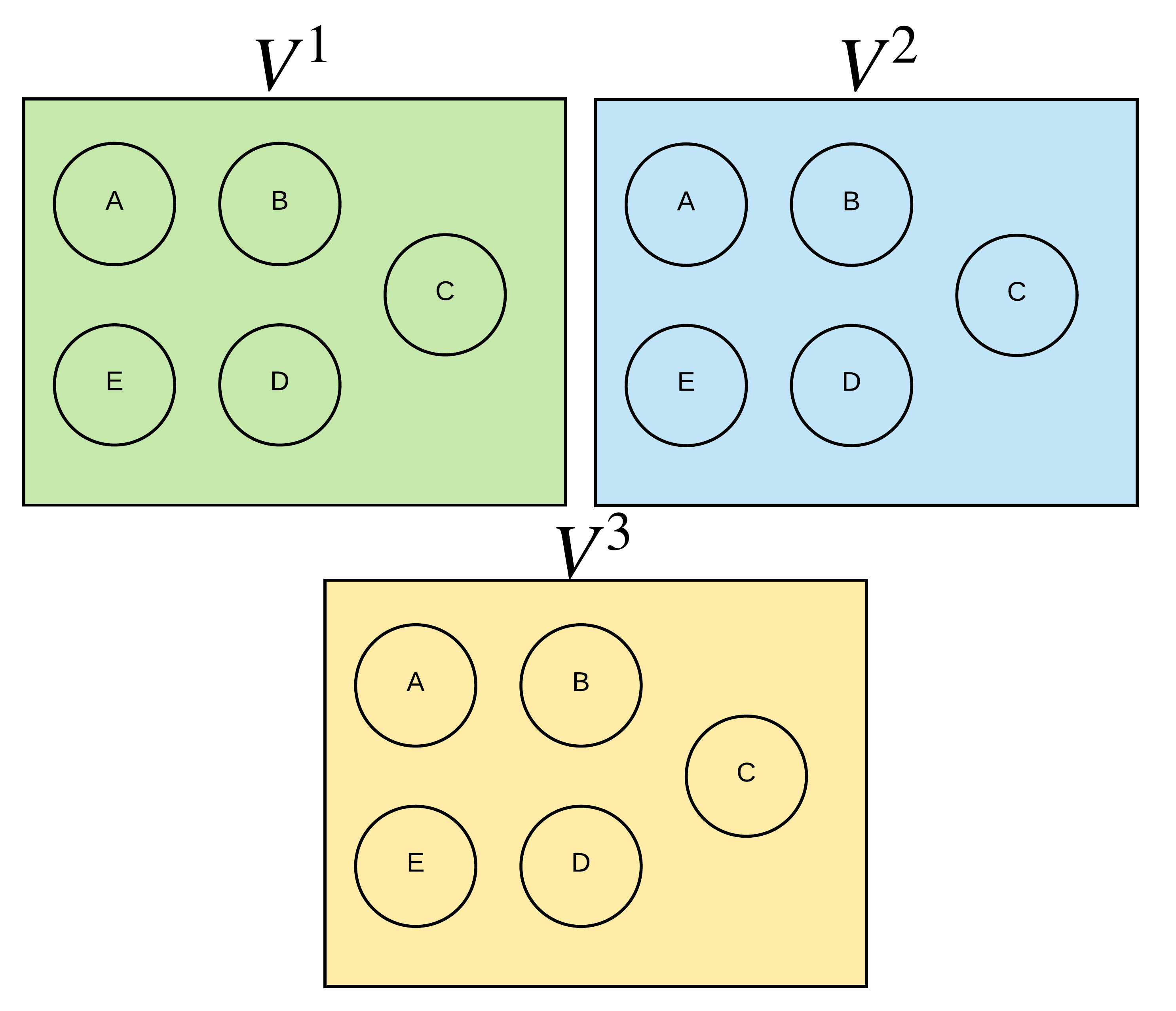}
         \caption{Initial three copies of $V$: $V^1$, $V^2$ and $V^3$.}
     \end{subfigure}
     \hfill
     \begin{subfigure}[b]{0.45\textwidth}
         \centering
         \includegraphics[scale=0.16]{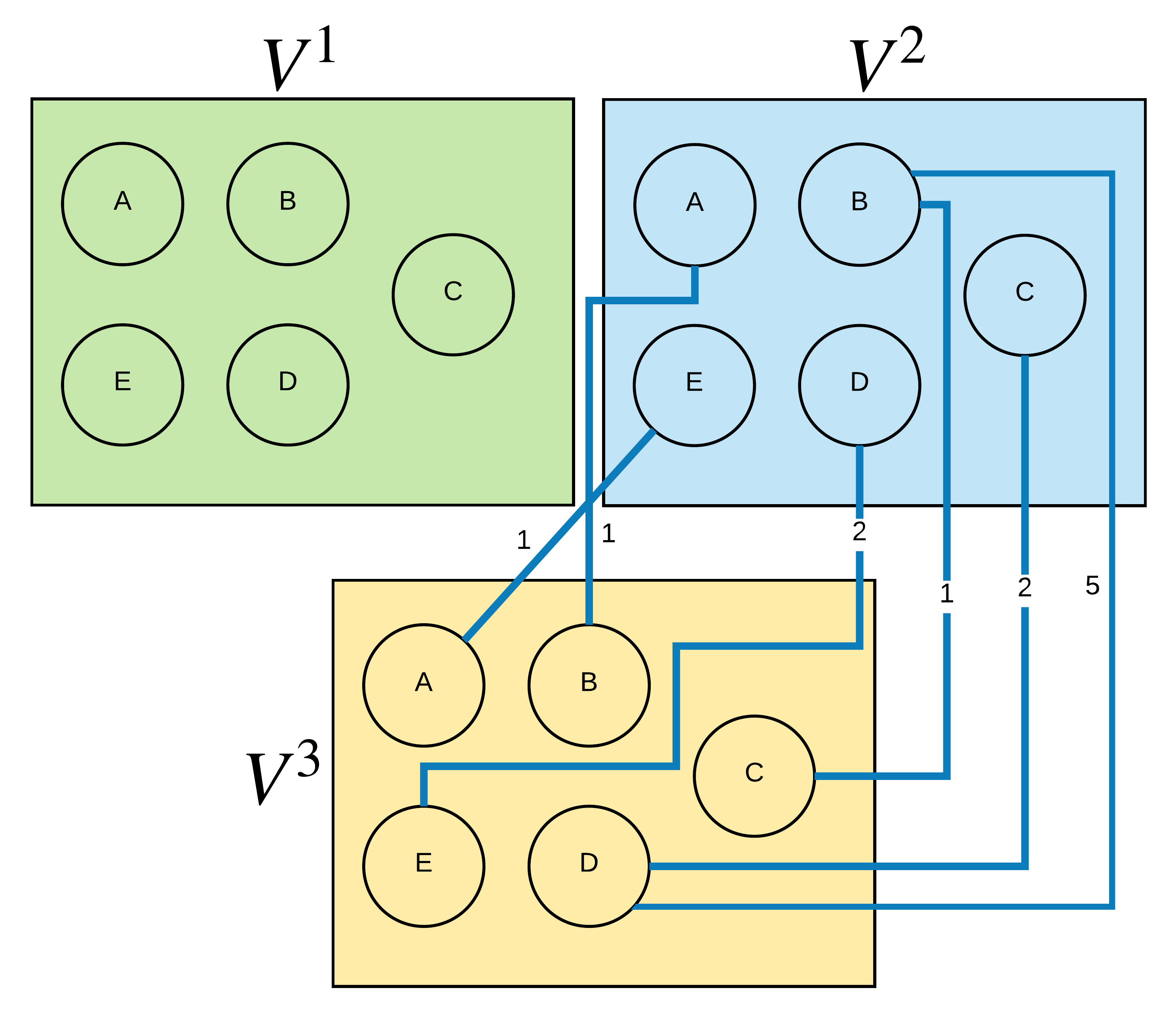}
         \caption{For every direct edge $(u,v) \in E_{G'}$, adding an edge from $u^2 \in V^2$ to $v^3 \in V^3$. }
     \end{subfigure}
     \vskip\baselineskip
     \begin{subfigure}[b]{0.45\textwidth}
     	\centering
     	\includegraphics[scale=0.16]{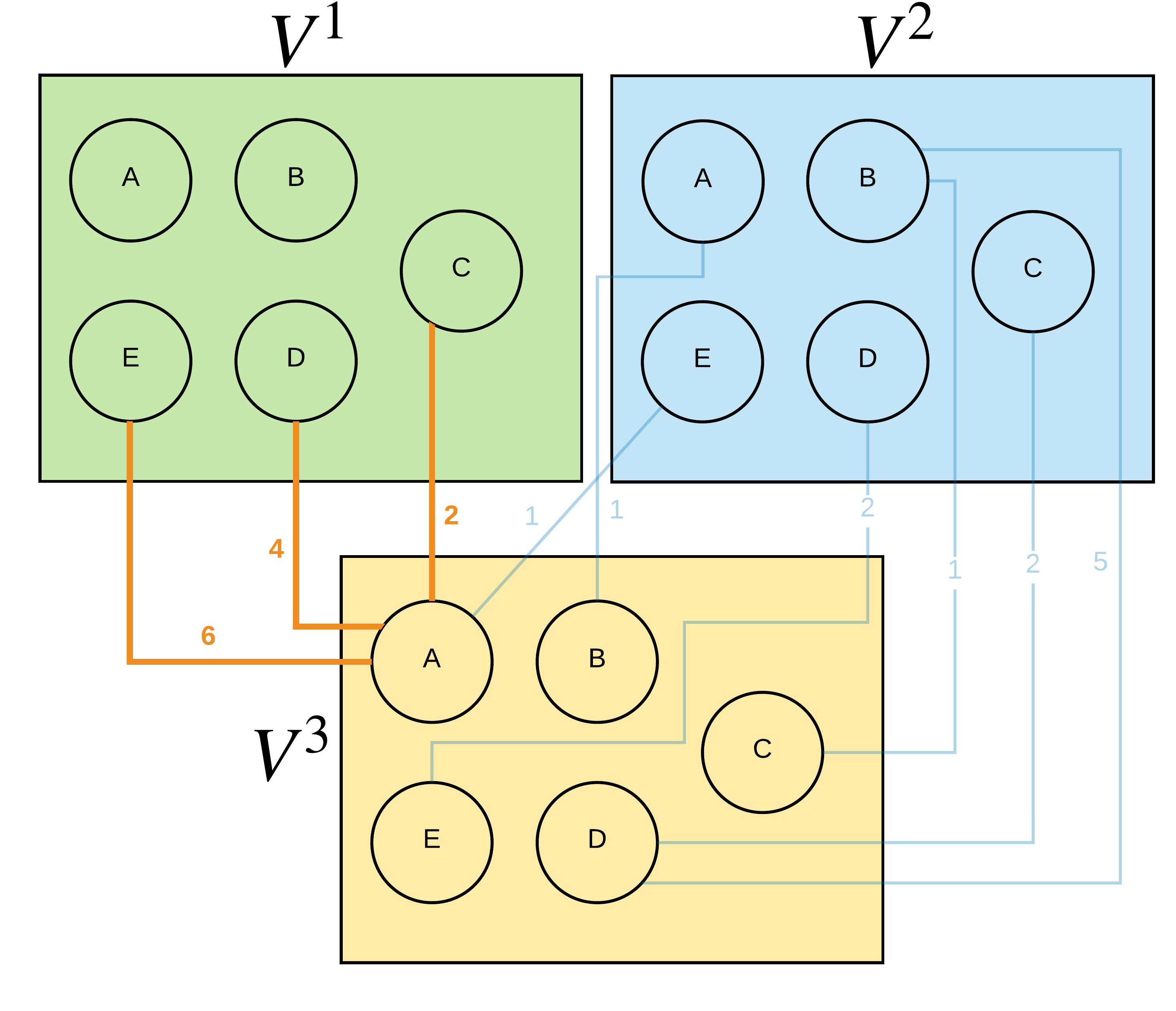}
      	\caption{Adding edges from $V^3$ to $V^1$ for which $D[u, v] < \infty$ (shown only for node A).}
     \end{subfigure}
     \quad
     \begin{subfigure}[b]{0.45\textwidth}
     	\centering
      	\includegraphics[scale=0.16]{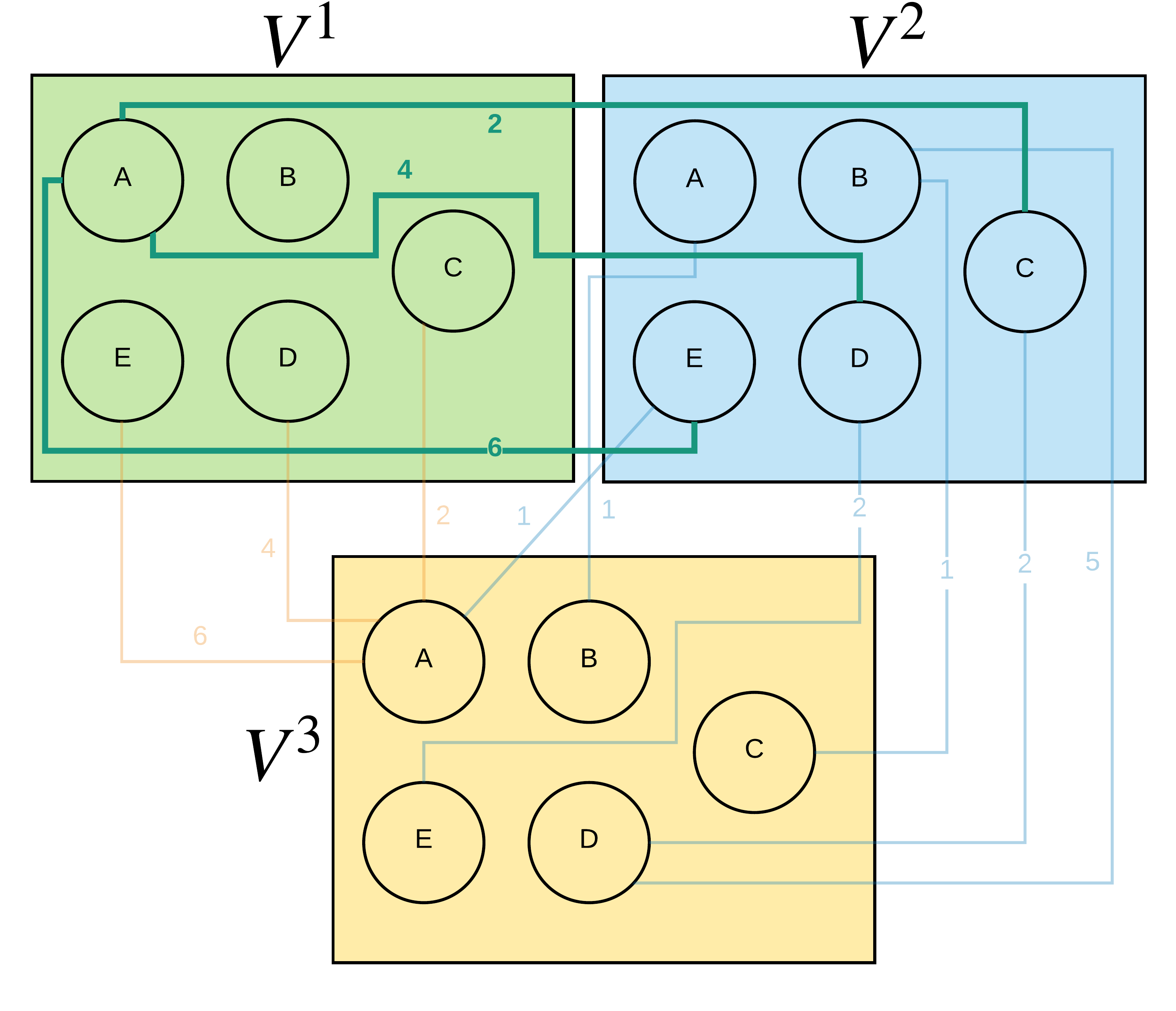}
      	\caption{Adding edges from $V^1$ to $V^2$ for which $D[u, v] < \infty$ (shown only for node A).}
     \end{subfigure}
    \caption{The process of finding minimum weight cycles as described in \Cref{sec:coreproces}.}
    \label{fig:g_prime_construction}     
\end{figure}

\noindent
Now that our problem is formulated as one of finding a cycle with a minimum sum of edge weights, we can turn to the task of actually finding this cycle. 
As noted earlier, finding the minimum weight cycle can be done by finding minimum weight triangles in undirected graphs \cite{Roditty2011MinimumAlgorithms}. 
This requires the following subproblems to be solved:

\begin{itemize}
	\item An auxiliary \textit{undirected} graph $G'$, comprised of three copies of the original graph, has to be created, as visualized in \Cref{fig:g_prime_construction}. Only the nodes are initially copied to $G'$, while the edges are constructed later (with specific rules that can be found in \cite{Roditty2011MinimumAlgorithms}). $E_G$ is then the set of edges in $G$ and $E_{G'}$ is the set of edges in $G'$. Similarly, the sets of nodes in $G$ and $G'$ are $V_G$ and $V_{G'}$ respectively. 
	\item The minimum triangle in $G'$ has to be found. The minimum triangle $C'$ has a direct correlation to the minimum weight cycle $C$ in $G$.
	\item A successor matrix (also called a \textit{witness matrix} in the literature) has to be calculated in order to find the nodes that make up $C$.
\end{itemize}

\subsection{Finding the minimum weight triangle}
\label{sec:coreproces}

\begin{figure}[t]
  \centering
  \includegraphics[width=.4\textwidth]{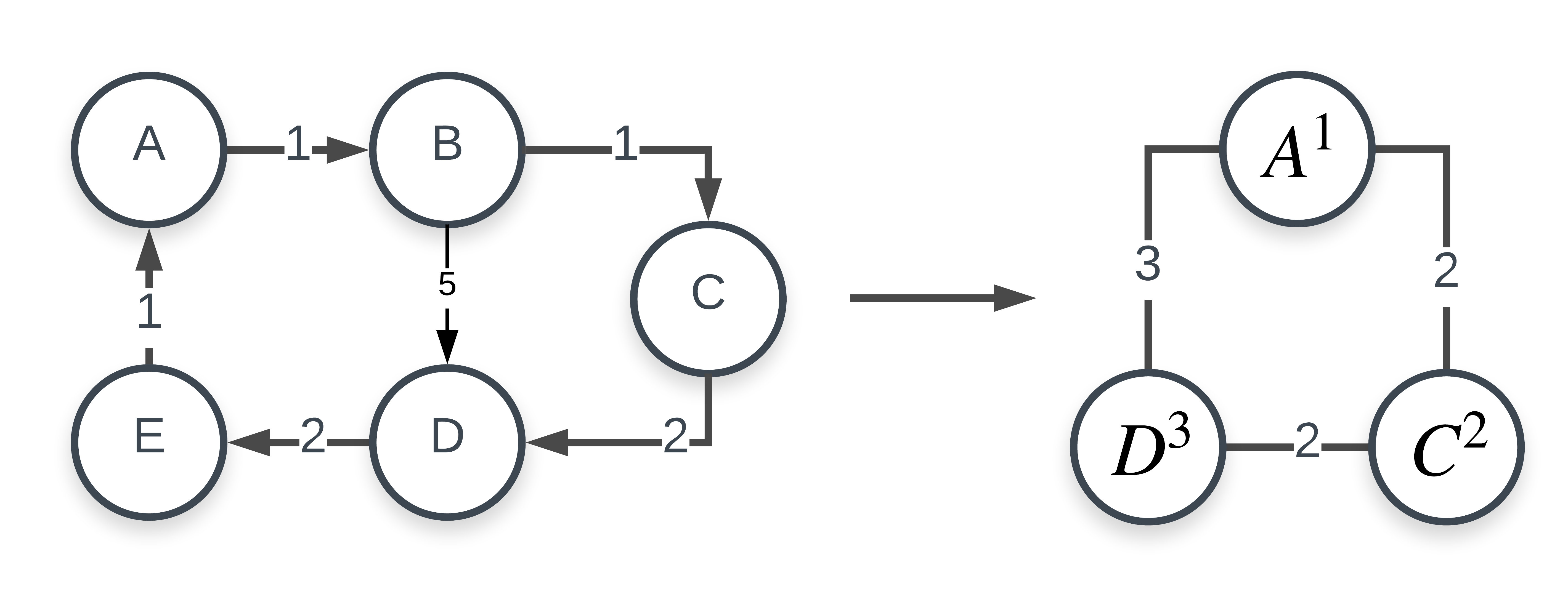}
  \caption{Example of a minimum weight triangle in $V_{G'}$ which has copies $V^1, V^2, V^3$. The superscript in the nodes denote which copy of $V$ the node is in.}
  \label{fig:min_weight_cycle}
\end{figure}

An example of converting a directed graph to an undirected minimum weight triangle is depicted in \Cref{fig:min_weight_cycle}. 
For ease of notation, when we write $u^x$, we refer to copy $x$ of a node $u \in V$ in set $V^x$. 
The edges in the triangle represent the sum of weights of the shortest path between nodes $u$ and $v$. 

In the algorithm, the first step is computing the $n \times n$ matrix $D$. 
The notation $D[u,v]$ for $u,v \in V$ is the computed weight distance between the nodes $u$ and $v$. $D[u,v] \geq d[u,v]$, where $d[u,v]$ is the minimum distance between $u$ and $v$.
Let $C = D \star D$ be the distance product for $1 \leq u,v \leq n$, as:
$$c_{uv} = \min_{k=1}^m \{a_{uk} + b_{vk} \}$$
Taking the distance product $D \star D$ gives the distance between nodes $u$ and $v$ in the $u, v$th entry with a number of steps $k \leq 2$. 
In order to find a minimum triangle, the following process is employed: 

\begin{itemize}
\item A tripartite graph $G'$ was constructed with a vertex set $V'$ (or $V_{G'}$) comprised of the partitions $V^1,V^2,V^3$, which were all copies of V. $G'$ would be undirected, even if the starting $G$ graph was directed. 
\item For every directed edge $(u,v) \in E_G$, an edge from $u^2 \in V^2$ to $v^3 \in V^3$ in $G'$ was constructed with weight $w(u,v)$.
\item $\forall u,v \in G$ such that $D[u,v] < \infty$, two edges were constructed in $G'$: one from $u^1 \in V^1$ to $v^2 \in V^2$ and another from $u^3 \in V^3$ to $v^1 \in V^1$, with weight $D[u,v]$.
\end{itemize}

\noindent
After this procedure, any triangle in $G'$ corresponds to a cycle in G. A triangle of nodes $(v_1, v_i, v_{i+1})$ was defined as the sum of the shortest path between $(v_1, v_i)$, the weight of $(v_i, v_{i+1})$ and the shortest path of $(v_{i+1}, v_1)$. 
In this triangle, the edge $(v_i, v_{i+1})$ was called the \textit{critical edge}.

\subsection{Calculating the witness matrix and finding the cycle}
\label{sec:stepback}

A witness matrix $\Pi '$ was defined so that, if $k= \Pi ' [u,v]$, then $k$ would be the successor of $i$ on a simple path from $u$ to $v$ with weight $D[u,v]$. 
By creating a witness matrix, the individual nodes that form the cycle found in G could be inferred. 
The formal definition of a witness matrix is the following: let D be an $n \times n$ distance matrix. 
An $n \times n$ matrix W is said to be a matrix of witnesses for the distance product $C = D \star D$ if for every $1 \leq u$, $v \leq n$ we have $1 \leq w_{uv} \leq n$ and $c_{uv} = d_{u,w_{uv}}+ d_{w_{uv},v}$.
Calculating a witness matrix is crucial to discover the actual nodes of the minimum weight cycle in $G$ after having found the minimum weight triangle in $G'$.
In fact, from the knowledge of the resulting three nodes in $G'$, it would be possible to retrieve the path of the cycle in $G$ by looking at the witness matrix.

An efficient way to calculate the witness matrix was introduced in \cite{Zwick2002AllMultiplication}.
First, it would be necessary to understand how to find a matrix of the \textit{unique witnesses}; as the name suggests, this matrix contains the information of the path for cycles with ``unique witnesses''. 
The algorithm to find the matrix of unique witnesses for a product $C'=D' \star D'$ is the following: for $1 \leq k \leq n$ and $1 \leq l \leq \ceil{ \log_2 n}+ 1$, $bit_l(k)$ is the $l-th$ bit in the binary representation of $k$ (k is a node of the graph); the $l-th$ bit is assumed to be the least significant.
For $1 \leq l \leq \ceil{\log_2 n}+ 1$, let $I_l = \{1 \leq k \leq n | bit_l(k) = 1\}$. 
The bit representation of the nodes is necessary to have a deterministic way to group them in different subsets $I_i$.

Let sampling be defined as the following: let $I \subset \{1, 2, \ldots, n\}$. $C_l = D[*, I_l] \star D[I_l, *]$, for $1 \leq l \leq \ceil{ \log_2 m} + 1$. 
Let $C_l = (c^{(l)}_{u,v})$. Finally, $c^{(l)}_{u,v} = c_{uv}'$ if there is a witness for $c_{uv}'$ whose l-th bit is 1; if this holds, then $w_{uv}$ would have the $l_{th}$ bit set to 1, and 0 otherwise. 
This means that the matrix of unique witnesses will be filled in $\ceil{ \log_2 m} + 1$ steps.

If there are elements in the network that have more than one witness, it is necessary to calculate a more general witness matrix: with very high probability, it is possible to find a witness matrix in $\log m$ steps. 
The algorithm is the following: for $1 \leq r \leq \log m$ steps, pick a random set of $s = c \log n$ subsets $R_t$ ($1 \leq t \leq s$), where every subset has size $\frac{m}{2^r}$. 
For each one of the subsets, find the matrix of unique witnesses for the product $D[*,R_s] \star D[R_s,*]$.
For each one of the unique witnesses $w_{uv}'$ found, check if it is also a witness for $C=D \star D$, or in other words, if $c_{uv} = D[u,w_{uv}'] \star D[w_{uv}',v]$; if it is, then $w_{uv}= w_{uv}'$. After all the cycles, the output of the algorithm is the final witness matrix W.
Then, it is then possible to backtrack the individual nodes of the minimum weight cycle in $G$ from the nodes of the minimum triangle in $G'$.

As a last step, in order to retrieve the true weight of a cycle $w(C)$, it is then necessary to look up the exchange rates in the lookup table and multiply all the exchange rates. 
In other words, after finding the nodes that would comprise the minimum weight cycle, the profit can be calculated by multiplying the weights of the edges in the original adjacency matrix. 
If the outcome value is $> 1$, it would mean that there is a mispricing that could be exploited.

\subsection{Complexity}
The minimum weight cycle algorithm as described in \Cref{sec:coreproces} runs in $\tilde{O}(Mn^\omega)$ time~\cite{Alon1997OnProblem}, where the $\tilde{O}$ symbol indicates that we do not take into account logarithmic factors, $M$ is the largest weight value in the network, and $\omega < 2.373$ is the smallest exponent of square matrix multiplication. 
Moreover, since the algorithm relies on matrix multiplication techniques, there is a relation between the cost of the algorithm and the largest value of the network (the product between large values is more expensive than the product between small values).
We note how the transformation of the weights in the network (\Cref{sec:weightmodi}) as well as the final translation back to the original data (\Cref{sec:stepback}), do not affect the overall complexity of the approach.


\section{Experiments} \label{Experiments}

This section explains first the experimental setup in \Cref{sec:expsetup}. 
After exploring the weight multiplication parameter in \Cref{sec:weightparam}, experiments were conducted to analyze the behavior of the algorithm on different subsets of the data. The experiments were conducted for various datasets, namely the dataset of all 16 markets in \Cref{ss:full_dataset}, one market with all currencies in  \Cref{sec:onemarket} and the five largest currencies in  \Cref{sec:fivelargst}. 
Note that we choose not to consider the trivial case of ``one currency, all markets''. 
Here, the optimal step would be to simply buy the currency on the market where the cost is lowest, and sell it on the market on which it has the highest value. 

\subsection{Experimental setup}
\label{sec:expsetup}

In this section, we apply the algorithmic approach outlined in \Cref{Our Approach} to the data described in \Cref{Datasets}.  
Two parameters were set: the weight multiplication parameter $c$, further discussed in \Cref{sec:weightparam} and the ``spread'' factor $\epsilon$. 
Whereas $\epsilon$ would in a real-world setting be dynamic and dependent on the considered currency, market and its volatility, we choose to set it to a random value in the range $\epsilon \in [0.99999,0.999999]$. 
As explained in \Cref{Problem statement}, 
this is a good approximation of the eventual spread factor in the currency conversion rate, as our data snapshot only contains the lowest ask price (lowest price you can buy for) and not the highest bid (the highest price you can sell for). 
Note how if no cycles with $p(C) \geq 1$ are found, non-simple cycles going back and forth between two nodes, could still be found. 
This happens when no cycle C reported a $p(C) \geq 1$.
Finally, we note that although currency names are preserved, we have chosen to anonymize the names of the cryptocurrency exchanges; the $16$ exchange markets are labeled $M_1$ to $M_{16}$.

\subsection{Weight multiplication parameter}
\label{sec:weightparam}

As explained in \Cref{sec:weightmodi}, the weight multiplier $c$ was used to map floating point numbers to integers. 
This parameter value had to be large enough to ensure that there would be as many unique integers as there are distinct floating numbers in the dataset. 
Since the percentage of transformed unique values given $c$ is directly dependent on the dataset, it is necessary to empirically test multiple values of $c$. 
A greater number of unique elements would make the algorithm more precise, although this would mean that the order of magnitude of the weights in the dataset would be greater.
The results can be found in Table~\ref{tab:parameter}. 
Since the algorithm's complexity depends on the size of the biggest number $M$ in the dataset, the trade-off is essentially between accuracy and speed.

\begin{table}[!t]
	\centering
	\begin{tabular}{l|ll}
		\textbf{c} &\textbf{Transformed unique} &\textbf{\% of weights} \\ \hline
		$10^2$ & 575 & 30.75\%  \\ 
		$10^3$ & 893 & 47.75\% \\ 
		$10^5$ & 1197 & 64.01\% \\ 
		$10^6$ & 1564 & 83.64\% \\ 
		$10^7$ & 1612 & 86.20\% 
	\end{tabular} 
	\caption{Unique edge weights (total: $1870$) after the transformation, for different values of $c$.}
	\label{tab:parameter}
\end{table}

The column titled ``Transformed unique'' refers to the number of unique integer edge weights present in the dataset after the transformation. 
Before the transformation, the values of the weights were in the interval $w \in [6.79^{-9}, 147291100]$, where $M = 147291100$. 
After the transformation, $w^{\rom{4}} \in [1, 376423]$, $M'= 376423$.

\subsection{Results --- All markets, all currencies} \label{ss:full_dataset}

The experiment was run three times; for each run, a different weight multiplier value was chosen. Specifically, $c \in \{ 10^2, 10^5, 10^7 \}$. These values were picked because the percentage of uniquely mapped values greatly differed, as can be seen in Table~\ref{tab:parameter}. The results of the three runs can be found in \Cref{tab:results_full}. Such a path should be interpreted as a sequence of actions. For example, the first path in \Cref{tab:results_full} can be interpreted as:

\begin{itemize}
    \item Sell JPY in $M_3$ (i.e. ``Market 3''), receive a common base currency.
    \item Buy JPY in $M_4$, using a common base currency.
    \item Buy BTC in $M_4$, using JPY.
    \item Buy IDR in $M_4$, using BTC.
    \item Buy BTC in $M_4$, using IDR.
    \item Buy JPY in $M_4$, using BTC.
    \item Sell JPY in $M_4$, receive a common base currency.
    \item Buy JPY in $M_3$, using a common base currency.
\end{itemize}

\noindent
Because we do not want to wait for transactions to complete, all trades should happen simultaneously. For this reason when exchanging currency between markets it is necessary to buy with and sell to a common base currency available in both markets, such as USD.

\begin{table}[!b]

  \begin{tabular}{lll|p{8.5cm}l}
    \textbf{$c$} &\textbf{|V|} &\textbf{|E|} &\textbf{Path} &\textbf{Profit}\\ \hline
     $10^2$ & 243 & 1718 & $M_3$/JPY, $M_4$/JPY, $M_4$/BTC, $M_4$/IDR, $M_4$/BTC, $M_4$/JPY, $M_3$/JPY & $-.003\%$ \\
    $10^5$ & 243 & 1718 & $M_3$/JPY, $M_4$/JPY, $M_4$/ETH, $M_4$/IDR, $M_4$/BTC, $M_4$/JPY, $M_3$/JPY & $+.343\%$ \\
    $10^7$ & 243 & 1718 & $M_3$/JPY, $M_4$/JPY, $M_4$/ETH, $M_4$/IDR, $M_4$/BTC, $M_4$/JPY, $M_3$/JPY & $+.343\%$ \\
  \end{tabular}
  \caption{Experimental results for the full data.}
  \label{tab:results_full}
\end{table}

Looking at the results in \Cref{tab:results_full}, it was clear that $c=10^2$ negatively affected the results because there were not enough unique integer weights in the transformed dataset. 
However, from $c \ge 10^5$, there was no more increase in profit.
It remains unclear which percentage threshold yields the optimal results; the ideal way to approach this problem would be to increase the value of $c$ until the runtime of the algorithm is greater than the response time of the servers of the cryptocurrency markets (or until the number of unique values is maximized).

The following experiments were run with the optimal value for $c$, i.e. the minimum value for which the percentage of unique transformed element was $100\%$; since the datasets were greatly decreased, $c=10^7$ was sufficient to reach optimality.

\subsection{Results --- One market, all currencies}
\label{sec:onemarket}

For this experiment, the algorithm was run on one market at a time, while trading across all available currencies. Three markets were chosen. These markets had very different sizes in terms of number of nodes and connections. The results can be seen in \Cref{tab:results_single_market}.

\begin{table}
  \begin{tabular}{lll|lll}
    \textbf{Market} &\textbf{|V|} &\textbf{|E|} &\textbf{Path} &\textbf{Profit}\\ \hline
    Market 1 & 194 & 516 & BTC, USD, ETH, BTC & $+.319\%$ \\
    Market 2 & 34 & 130 & BTC, SC, ETH, BTC &$-.0004\%$ \\
    Market 3 & 21 & 92 & ETC, USD, BCH, USD, ETC & $-.00002\%$ \\	
    Five largest &  51 & 590 & $M_5$/ETH, $M_5$/BTC, $M_6$/BTC, $M_7$/BTC, $M_7$/ETH, $M_5$/ETH & $+.359\%$ \\

  \end{tabular}
  \caption{Experimental results for three single markets and all currencies as well as all markets and the five largest currencies.}
  \label{tab:results_single_market}
\end{table}

Without having the possibility of trading between various financial markets, it was clear that sometimes there are not enough mispricings to make a profit out of a cycle. In two out of three of these runs the profit was negative, which meant that following the minimum weight cycle would result in a loss; any other cycle would produce a greater loss, and no cycle would provide any profit.

\subsection{Results -- All markets, five biggest currencies}
\label{sec:fivelargst}

For the last experiments, the 5 cryptocurrencies with the largest market capitalization were chosen, namely: Bitcoin, Ethereum, Ripple, Bitcoin Cash and Litecoin.
This test was conducted to see if ignoring the small cryptocurrencies would also end up in a profit, and if the profit was comparable to the one achieved in the full data. 
The result can be seen in the bottom line of \Cref{tab:results_single_market}. 
The found profit is the largest profit found over all experiments. 
Why this cycle was not found in the full dataset could be explained by the fact that not all weights are mapped to unique integers. 
Using $c = 10^5$, this subgraph transformed 91.92\% of the values to unique integers, whereas the full dataset only transformed 64.01\% to unique integers as shown in \Cref{tab:parameter}. 
For $c = 10^7$, 86.2\% of the weight values mapped to unique integers, which is still lower than the value achieved on this smaller subgraph. 
Indeed, setting $c = 10^7$ for this subgraph yields a 100\% unique weight mapping but does not yield a different result.


\section{Conclusion} \label{Conclusions}

The goal of the paper was to find a minimum weight cycle in a graph representing a market of cryptocurrencies that are being exchanged at different marketplaces. 
Such a cycle potentially indicates a mispricing between the currencies, allowing a profit to be made by executing the sequence of trades modeled by the discovered cycle. 
To perform this task efficiently, we successfully transformed a real-world dataset of $110$ currencies being traded at $16$ cryptocurrency marketplaces into a suitable graph dataset. 
For this, the edge weights representing exchange rates were modified in order to formulate the problem in terms of a minimization of sums instead of a maximization of products (the exchange rates). 
This involved an edge weight multiplier parameter, for which several experiments with different parameter values were performed. 
As this is essentially a heuristic step, the conversion of weights could potentially result in a graph structure in which a profitable cycle is overlooked. 
However, this was not problematic as we can simply observe that the obtained cycle would not result in a profit, ignore the finding and not execute the trades. 
Results demonstrate how indeed cycles with a net profit can be found in this type of cryptocurrency market network data. 
Moreover, it was not only mispricings between currencies at one marketplace, but actually also differences between multiple marketplaces that showed a significant profit.

In future work, we would like to improve the quality and performance of the implementation such that it can efficiently be applied in practice.
In particular, performance optimization may need to be done to reduce the wall clock time such that the computation can be done quickly enough so that a discovered cycle's trades can actually take place, before the prices have changed. 
To achieve this, the algorithm could potentially be parallelized with regard to computing the witness matrix: its computation requires a multitude of independently computed unique witnesses, which can be sped up by a parallelized implementation.
In addition, we want to investigate the effect of real-world aspects such as trading fees and available volumes to assess to what extent the actual profit ratios can also result in significant profit volume on real-world cryptocurrency marketplaces.


\subparagraph*{Acknowledgements.}

The authors thank Cristian Mesaros of T.K.Boesen Capital Partners ApS for  ideas, useful comments and suggestions. 
The third author is supported by funding from the European Research Council (ERC) under the EU Horizon 2020 research and innovation programme (grant agreement 638946).


\bibliography{lipics-v2016-sample-article}

\end{document}